\documentclass[10pt,journal]{IEEEtran}
\usepackage{amsmath,graphicx}
\usepackage[T1]{fontenc}
\usepackage{amsmath}
\usepackage{amsfonts}
\usepackage{amssymb}
\usepackage{bm}
\usepackage[noadjust]{cite}
\usepackage[nolist,printonlyused]{acronym}
\usepackage{algorithm}
\usepackage{algpseudocode}
\usepackage{mathtools}

\usepackage{blindtext}

\renewcommand{\baselinestretch}{1.0}

\DeclareFontFamily{U}{mathx}{\hyphenchar\font45}
\DeclareFontShape{U}{mathx}{m}{n}{
	<5> <6> <7> <8> <9> <10>
	<10.95> <12> <14.4> <17.28> <20.74> <24.88>
	mathx10
}{}
\DeclareSymbolFont{mathx}{U}{mathx}{m}{n}
\DeclareFontSubstitution{U}{mathx}{m}{n}
\DeclareMathAccent{\widebar}{0}{mathx}{"73}

\usepackage{tikz}
\usepackage{pgfplots}
\pgfplotsset{compat=newest}
\usetikzlibrary{spy}

\usepackage{times}

\usepackage{glossaries}
\newacronym{ris}{RIS}{Reconfigurable intelligent surface}

\title{Compressed Sensing Constant Modulus Constrained Projection Matrix Design and High-Resolution DoA Estimation Methods}

\author{Khaled Ardah and Martin Haardt
	\thanks{{The authors gratefully acknowledge the support of the German Research Foundation (DFG) under contract no.~HA 2239/6-2 (EXPRESS II).}}
	\thanks{K. Ardah and M. Haardt are with Communications Research Laboratory (CRL), TU Ilmenau, Ilmenau, Germany (e-mail: \{khaled.ardah, martin.haardt\}@tu-ilmenau.de).}}

\def\kr#1{\textcolor{black}{#1}}

\begin{document}
\maketitle

\begin{abstract}
This paper proposes a compressed sensing-based high-resolution direction-of-arrival estimation method called gradient orthogonal matching pursuit (GOMP). It contains two main steps: a sparse coding approximation step using the well-known OMP method and a sequential iterative refinement step using a newly proposed gradient-descent method. To enhance the recoverability, we further propose an efficient projection matrix design method, which considers the constant modulus constraints imposed by the projection matrix hardware components. Simulation results show the effectiveness of the proposed methods as compared to benchmark \kr{algorithms}.
\end{abstract}
\vspace{-5pt}

\vspace{-12pt}
\section{Introduction}
Direction-of-arrival (DoA) estimation is a fundamental problem in a large variety of applications including radar and wireless communications. Over the years, various DoA estimation methods have been proposed \cite{MLDoA}, including the subspace-based methods MUSIC and ESPRIT \cite{MUSIC,StESBRIT}. While subspace-based methods have been shown to perform asymptotically optimal, these methods suffer from a performance degradation in difficult scenarios such as high noise power and low number of measurements. Recently, compressed sensing (CS) techniques \cite{CS} have been introduced as an attractive alternative to subspace-based methods. By exploiting the sparsity property, it has been shown that CS-based techniques exhibit a good estimation accuracy, even in difficult scenarios and are able to recover the signal of interest from far fewer samples than required by traditional acquisition systems \cite{CSBounds,ardah2021recovery}. 


Generally, CS-based methods can be categorized into on-grid, off-grid, and gridless \cite{Boche2015}. The former approximates the DoA parameters using a given dictionary, which is formed by sampling the desired domain into a number of grid points. A well-known on-grid CS-based technique is the orthogonal matching pursuit (OMP) \cite{OMPComp}. However, if the true DoAs belong to a continuous domain, which is the case in general, on-grid methods suffer from the sampling errors, since the number of grid-points are practically limited. To overcome this issue, off-grid methods extend the on-grid methods by introducing a refinement step to reduce or even eliminate the sampling errors \cite{NOMP}. \kr{On the other hand}, gridless  methods \cite{ardah_icassp19} do not require a dictionary, but rather operate directly on the continuous domain, thus, avoiding sampling errors. However, gridless methods are, generally, very complex especially with large-scale arrays, which limits their practical implementation. 

In this paper, as our first contribution, we propose a high-resolution CS-based DoA estimation method called Gradient OMP (GOMP). Specifically, GOMP contains two steps. In Step I, the OMP method is used to approximate the true DoA parameters to their best on grid points. After that, the approximated DoA parameters are refined sequentially and iteratively in Step II using a novel proposed method to eliminate the sampling errors. Simulation results show that GOMP exhibits a high-resolution DoA estimation accuracy with low computational complexity. 

In CS-based methods, the sensing matrix should be designed carefully so that it satisfies a certain property, e.g., the mutual coherence to enhance recoverability of desired signal \cite{CS,ardah2021recovery,CSBounds,ardah_icassp2020}. In general, the sensing matrix comprises a tunable $N\times M$ projection matrix and a fixed $M\times P$ dictionary matrix, where $N$, $M$, and $P$, with $N\leq M \leq P$, denote the number of dedicated radio-frequency (RF) chains, the number of array elements, and the number of sampling (grid) points, respectively. Considering such a structure, several sensing matrix design methods via mutual coherence minimization have been proposed, e.g., in \cite{CSSMCM,GDes,GDes2,Direct,SVDShrinkage,ardah2021recovery}. Assuming that $M = P$, we have proposed \kr{in \cite{CSSMCM}} a sensing matrix design method called sequential mutual coherence minimization (SMCM), where the original nonconvex problem is relaxed and divided into several convex sub-problems that are updated sequentially using alternating optimization. On the other hand, gradient-descent (GD)-based sensing matrix design methods are proposed in \cite{GDes} and \cite{GDes2}. In contrast to \cite{GDes}, the method in \cite{GDes2} imposes the unit-norm constraints directly into the objective function, leading to a better solution. 

In many applications, however, the projection matrix is realized using, e.g., a network of tunable phase-shifters \cite{ardah_unify}. This implies that the entries of the projection matrix are subject to constant modulus (CM) constraints so that only their phase angles are adjustable. In \cite{ardah2021recovery}, we have proposed a GD-based projection matrix design method, which takes into account, for the first time, the CM constraints imposed by the projection matrix hardware components. In this paper, as our second contribution, we propose a new GD-based projection matrix design method by extending and enhancing the proposed methods in \cite{ardah2021recovery} and \cite{GDes2} while taking into account the CM constraints. Using computer simulations, we show that the new proposed GD-based method outperforms the baseline methods leading to a higher DoA estimation accuracy.



\section{System model}

We assume that $K$ narrow-band far-field source signals impinge on an array composed of $M$ omni-directional sensors. The output vector at the $\ell$th time sample is given by 
\begin{align}\label{ybar}
      \bar{{\bf y}}_{[\ell]} = {\bf A}(\bm{\nu}) {\bf x}_{[\ell]} + {\bf \bar n}_{[\ell]} \in \mathbb{C}^{M}, \quad \ell \in \{1,\dots,L\},
\end{align}
where ${\bf x}_{[\ell]} \in \mathbb{C}^{K}$ is the vector of waveforms emitted by the $K$ sources, ${\bf \bar n}_{[\ell]} \in \mathbb{C}^{M}$ contains the spatially and temporally white circular Gaussian sensor noise, and $L$ denotes the number of available time samples. The matrix ${\bf A}(\bm{\nu}) = [{\bf a}(\nu_1), \dots, {\bf a}(\nu_K)] \in \mathbb{C}^{M\times K}$ denotes the true array steering matrix, whose $k$th column is the array response vector ${\bf a}(\nu_k)$ corresponding to the DoA of the $k$th source $\nu_k$, $k \in \{1,\dots,K\}$. Here, $ \nu_k = \frac{2\pi d}{\lambda}  \sin(\theta_k)$ is the DoA parameter in the spatial frequency domain, $\theta_k \in [-90^{\circ},90^{\circ}]$ is the DoA parameter in the angular domain, $d$ is the inter-sensor spacing, $\lambda$ is the wavelength.  We denote $\bm{\nu} = [\nu_1,\dots,\nu_K]^{\mathsf{T}} \in \mathbb{R}^{K}$ as the true DoA parameter vector. The array response vector ${\bf a}(\nu)$ describes a manifold denoted as $\mathbb{M}^{M}$. \kr{In this paper, we assume} a uniform linear array (ULA), where ${\bf a}(\nu)$ is given by $	{\bf a}(\nu) = [1,e^{j \nu },\dots,e^{j(M-1) \nu}]^{\mathsf{T}}  \in \mathbb{C}^{M}$.   

Let $\bm{\Phi} \in \mathbb{C}^{N\times M}$ denote the analog projection matrix, which compresses the output of $M$ antennas to $N$ active RF channels. We assume that the analog domain is realized in hardware using a network of phase shifters. Therefore, the entries of $\bm{\Phi}$ must satisfy the CM constraints as $|\phi_{n,m}| = 1, \forall n,m$, where $\phi_{n,m}$ is the $(n,m)$th entry of $\bm{\Phi}$. Then, the complex baseband array output in (\ref{ybar}), after projection, can be expressed as  
\begin{align}\label{y}
	{{\bf y}}_{[\ell]} = \bm{\Phi} \bar{{\bf y}}_{[\ell]} = \bm{\Phi} {\bf A}(\bm{\nu}) {\bf x}_{[\ell]} + {\bf n}_{[\ell]} \in \mathbb{C}^{N},
\end{align}
where ${\bf n}_{[\ell]} =  \bm{\Phi} {\bf \bar n}_{[\ell]} \in \mathbb{C}^{N}$. Let ${\bf X} = [{\bf x}_{[1]},\dots,{\bf x}_{[L]}]  \in \mathbb{C}^{K \times L}$ and  ${\bf N} = [{\bf n}_{[1]},\dots,{\bf n}_{[L]}]  \in \mathbb{C}^{N \times L}$. Then, the measurement matrix ${\bf Y}  = [{\bf y}_{[1]},\dots,{\bf y}_{[L]}] \in \mathbb{C}^{N\times L}$ can be written as 
\begin{align}\label{Y}
{\bf Y} =  \bm{\Phi} {\bf A}(\bm{\nu}) {\bf X} + {\bf N} \in \mathbb{C}^{N\times L}.
\end{align}

In Section \ref{sec3}, we propose the Gradient OMP (GOMP) method for estimating the DoA parameter vector $\bm{\nu}$ and the signal matrix ${\bf X}$ from ${\bf Y}$ with \kr{a} high-resolution. Next, we propose in Section \ref{sec4} a new GD-based projection matrix design method while taking into account the CM constraints imposed by the projection matrix hardware components.



\section{Proposed GOMP For DoA estimation}\label{sec3}

\subsection{The single-source case}

To simplify the exposition, we first assume a single-source case, i.e., $K = 1$ and that $ \bm{\Phi}$ is given and fixed. Therefore, after removing the source index $k$, (\ref{Y}) simplifies to
\begin{align}\label{sig}
{\bf Y} = \bm{\Phi} {\bf a}(\nu) {\bf x}^{\mathsf{T}} +  {\bf N} \in \mathbb{C}^{N\times L},  
\end{align}
where ${\bf x} \in \mathbb{C}^{L}$. To estimate the spatial frequency ${\nu}$ and the source signal ${\bf x}$, the problem can be written as 
\begin{align}\label{OPT11}
\underset{{\bf x}, \nu}{\arg\min}  \Vert {\bf Y} - \bm{\Phi} {{\bf a}}({\nu}) {\bf x}^{\mathsf{T}} \Vert^{2}_{\text{F}},
\end{align} 
which is non-convex due to its joint optimization. Therefore, we decouple the recovery procedure between ${\nu}$ and ${\bf x}$, where we update for one variable assuming the other is fixed. 

First, its not hard to see that the optimal solution to ${\bf x}$, in the least squares sense, for a fixed $\nu$ is given as 
\begin{align}\label{xLs}
{\bf x} = (\{ \bm{\Phi}{{\bf a}}({\nu})\}^{+}{\bf Y})^{\mathsf{T}} \in \mathbb{C}^{L}.
\end{align}  

On the other hand, to estimate $\nu$ for a fixed ${\bf x}$, we start by writing the steering vector ${\bf a}(\nu)$ using its first order Taylor series expansion as 
\begin{align}\label{aTaylor}
{\bf a}(\nu) \approx {\bf a}(\mathring{\nu}) + {\bf g}(\mathring{\nu})  \cdot  \delta \in \mathbb{C}^{M}, 
\end{align}
where $\mathring{\nu}$ is a known close approximation to $\nu$ (e.g., the best on-grid sampling frequency), $\delta = \nu - \mathring{\nu}$ is the sampling error between the true $\nu$ and the approximated $\mathring{\nu}$, and 
${\bf g}(\mathring{\nu})$ is the gradient vector of ${\bf a}(\mathring{\nu})$ with respect to $\mathring{\nu}$ which is given as
\begin{align}
{\bf g}(\mathring{\nu}) = j \text{diag}\{0,1,\dots,M-1\} {\bf a}(\mathring{\nu}) \in \mathbb{C}^{M}.
\end{align}

By substituting (\ref{aTaylor}) into (\ref{sig}), the matrix ${\bf Y}$ can be written as  	
\begin{align}\label{sig2}
{\bf Y} \approx \bm{\Phi} {\bf a}(\mathring{\nu}) {\bf x}^{\mathsf{T}} + \bm{\Phi} {\bf g}(\mathring{\nu})  {\bf x}^{\mathsf{T}} \cdot \delta  +  {\bf N},
\end{align}		
where the only unknown variable is the sampling error $\delta$. To this end, we note from (\ref{aTaylor}) that for any given $\mathring{\nu}$, the true DoA spatial frequency $\nu$ can also be obtained as\kr{
\begin{align}\label{mutrue}
\nu = \mathring{\nu} + \delta,
\end{align}}
where $\delta$ here can be seen as the step-size that indicates the shift value (i.e., the sampling error correction) that should be added to or subtracted from $\mathring{\nu}$, depending on its sign, so that we have a perfect estimation of $\nu$. 
Therefore, for a given approximation $\mathring{\nu}$, finding the true DoA parameter $\nu$ requires finding the best step-size $\delta$. The problem can be written as 
\begin{align}\label{deltaP}
\underset{\delta}{\arg\min} \Vert {\bf Y} - \bm{\Phi} {\bf a}(\mathring{\nu}) {\bf x}^{\mathsf{T}} - \bm{\Phi} {\bf g}(\mathring{\nu})  {\bf x}^{\mathsf{T}} \cdot \delta  \Vert_{\text{F}}^2.
\end{align}

By applying the $\text{vec}\{ {\bf A}  \text{diag}\{ {\bf b}\} {\bf C}\} = ({\bf C}^{\mathsf{T}} \diamond {\bf A}) {\bf b}$ property, where $\diamond$ denotes the Khatri-Rao product, the vectorized form of ${\bf Y}$ can be expressed as
\begin{align}\label{sig22}
	{\bf y} = \text{vec}\{{\bf Y}\} = ({\bf x} \diamond \bm{\Phi} {\bf a}(\mathring{\nu}))  + ({\bf x} \diamond \bm{\Phi} {\bf g}(\mathring{\nu})) \cdot \delta +  {{\bf n}},
\end{align}
where ${{\bf n}} = \text{vec}\{{\bf N} \}$. Therefore, (\ref{deltaP}) can be rewritten as 
\begin{align}\label{deltaP2}
	\underset{\delta}{\arg\min} \Vert {\bf y} - ({\bf x} \diamond \bm{\Phi} {\bf a}(\mathring{\nu}))  - ({\bf x} \diamond \bm{\Phi} {\bf g}(\mathring{\nu})) \cdot \delta  \Vert_{2}^2,
\end{align}
so that \kr{a real-valued solution to $\delta$,} for fixed $\mathring{\nu}$ and ${\bf x}$, can be obtained as
\begin{align}\label{delta}
\delta =  \Re \big\{  \{{\bf x} \diamond \bm{\Phi} {\bf g}(\mathring{\nu}) \}^{+} ({\bf y} - ({\bf x} \diamond \bm{\Phi} {\bf a}(\mathring{\nu})) ) \big\}.
\end{align}					

Using (\ref{xLs}), (\ref{mutrue}) and (\ref{delta}), a solution to (\ref{OPT11}), i.e., to ${\bf x}$ and to $\nu$ can be obtained as \kr{follows}: Let us denote $\mathring{\nu}$ by ${\nu}^{(0)}$, i.e., the first estimate of $\nu$ obtained using \kr{a CS-based on-grid method}, e.g. OMP. Then, the first estimate of ${\bf x}$, i.e., ${\bf x}^{(0)}$ can be obtained using (\ref{xLs}). After that, we refine $\delta^{(i)}$, $\nu^{(i)}$, and ${\bf x}^{(i)}$ iteratively, as summarized in Algorithm~\ref{GOMP}, where $i$ denotes the \textit{inner} iteration index. The proposed iterative steps are repeated until a maximum number of iterations $I_{\text{max}}$ is reached. Note that, in Step \ref{UAC}, we have included an \textit{update acceptance condition}, which implies that GOMP accepts the new updates ${\nu^{(i+1)}}$ and ${{\bf x}^{(i+1)}}$ only if they \kr{reduce} the current cost-function $\epsilon^{(i)}$. This means that Algorithm~\ref{GOMP} has a monotonic and a guaranteed convergence, since we always have $\epsilon_{(i+1)} \leq \epsilon_{(i)}$ and the cost-function $\epsilon$ is bounded as $0\leq \epsilon_{(i)} \leq \epsilon_{(0)}$.

\begin{algorithm}[t]
	\caption{GOMP: The single-source case.}
	\label{GOMP}
	\begin{algorithmic}[1]
		\State{\textbf{Input:} Measurement matrix ${\bf Y}$ and initials ${\nu}^{(0)}$ and ${\bf x}^{(0)}$}
		\State{Compute $\epsilon^{(0)} = \Vert{\bf Y} - \bm{\Phi}{\bf a}(\nu^{(0)}) ({\bf x}^{(0)})^{\mathsf{T}}  \Vert^{2}_{\text{F}}$}
		\State{Choose $I_{\text{max}}$ and set $i = 1$ }
		\While{ $i \leq I_{\text{max}}$}
		\State{For given ${\bf x}^{(i-1)}$ and ${\nu}^{(i-1)}$, compute ${\delta}^{(i)}$  using (\ref{delta})}
		\State{For given ${\nu}^{(i-1)}$ and $\delta^{(i)}$, compute $\nu^{(i)}$ using (\ref{mutrue})}
		\State{For given $\nu^{(i)}$, compute ${\bf x}^{(i)}$ using (\ref{xLs})}
		\State{Compute $\epsilon^{(i)} = \Vert{\bf Y} -  \bm{\Phi}{\bf a}(\nu^{(i)}) ({\bf x}^{(i)})^{\mathsf{T}}  \Vert^{2}_{\text{F}}$}
		\If{$\epsilon^{(i)} > \epsilon^{(i-1)}$}\label{UAC}
		\State{Return ${\hat \nu} = {\nu}^{(i-1)}$ and ${\bf \hat x} = {\bf x}^{(i-1)}$ and terminate}
		\EndIf
		\EndWhile
		\State{\textbf{Output:} The estimated parameters ${\hat \nu}$ and ${ \bf \hat x}$}
	\end{algorithmic}
\end{algorithm}

\subsection{The general case}
For the general case, i.e., $K \geq 1$, the signal model in (\ref{Y}), i.e., ${\bf Y} =  \bm{\Phi} {\bf A}(\bm{\nu}) {\bf X} + {\bf N}$ can also be written as 
\begin{align}
{\bf Y} = \bm{\Phi}  {\bf a}({\nu_k}) {\bf x}^{\mathsf{T}}_k + \sum_{k^{\prime} \neq k}^K \bm{\Phi}  {\bf a}({\nu_{k^{\prime}}}) {\bf x}^{\mathsf{T}}_{k^{\prime}}  + {\bf N} \in \mathbb{C}^{N\times L},
\end{align} 
where ${\bf x}_k \in \mathbb{C}^{L}$ is the $k$th column-vector of ${\bf X}^{\mathsf{T}} \in \mathbb{C}^{L \times K}$ and ${\bf a}({\nu_k}) \in \mathbb{C}^{M}$ is the $k$th column vector of  ${\bf A}(\bm{\nu})$. Therefore, Algorithm \ref{GOMP} can be utilized to estimate ${\nu}_k$ and ${\bf x}_k, \forall k$, sequentially as summarized in Algorithm~\ref{GOMP2}. In Step~\ref{omps}, for a given measurement matrix ${\bf Y}$ and a dictionary matrix ${\bf \mathring A }(\bm{\mathring \nu}) \in \mathbb{C}^{M \times P}$ with $P \gg K$, we use the CS-based on-grid method OMP to obtain the initial estimates\footnote{Note that (\ref{Y}) can be written in a sparse-form as \cite{ardah_icassp2020}: ${\bf Y} \approx  \bm{\Phi} {\bf \mathring A}(\bm{\mathring \nu}) {\bf \mathring X} + {\bf N} \in \mathbb{C}^{N\times L}$, where ${\bf \mathring A} (\bm{\mathring \nu}) = [{\bf a}(\mathring \nu_1), \dots, {\bf a}(\mathring\nu_P)] \in \mathbb{C}^{M\times P}$ denotes a dictionary matrix formed using the grid points $\bm{\mathring \nu} = [\mathring \nu_1,\dots, \mathring \nu_P]^{\mathsf{T}} \in \mathbb{R}^{P}$, with $P \gg K$ denoting the number of grid points, and ${\bf \mathring X} \in \mathbb{C}^{P \times N}$ is an ${K}$ row-sparse matrix. Therefore, the on-grid CS-based techniques, e.g., OMP \cite{OMPComp} can readily be applied to estimate $\bm{\nu}^{(0)}$ and ${\bf  X}^{(0)}$.} of the spatial frequency vector $\bm{\nu}^{(0)}  \in \mathbb{C}^{K}$ and the source signal matrix ${\bf X}^{(0)}\in \mathbb{C}^{K \times L}$. After that, Algorithm~\ref{GOMP2} runs for a maximum of $J_{\text{max}}$ \textit{outer} iterations to refine the initial parameters ${\nu}^{(0)}_k$ and ${\bf x}^{(0)}_k, \forall k$, sequentially. Specifically, to refine the $k$th source parameters, i.e., ${\nu}^{(j-1)}_k $, and $ {{\bf x}}^{(j-1)}_k$ at the $j$th \textit{outer} iteration, we first calculate ${\bf Y}^{(j)}_k$ as shown in Step~\ref{stepYk}, where ${\bf Y}^{(j)}_{\text{new},k}$ is calculated using the updated parameters in the current $j$th iteration, i.e., $\nu^{(j)}_{1}, \dots, \nu^{(j)}_{k-1} $ and ${\bf x}^{(j)}_{1}, \dots, {\bf x}^{(j)}_{k-1}$ while ${\bf Y}^{(j)}_{\text{old},k}$ is calculated using the updated parameters in the previous $(j-1)$th iteration, i.e., $\nu^{(j-1)}_{k+1}, \dots, \nu^{(j-1)}_{K} $ and ${\bf x}^{(j-1)}_{k+1}, \dots, {\bf x}^{(j-1)}_{K}$. Given ${\bf Y}^{(j)}_k$, ${\nu}^{(j-1)}_k $, and $ {{\bf x}}^{(j-1)}_k$, Algorithm~\ref{GOMP} is then used in Step \ref{stepGOMP} to obtain ${\nu}^{(j)}_k $ and $ {{\bf x}}^{(j)}_k$. 

\begin{algorithm}[t]
	\caption{GOMP: The general case.}
	\label{GOMP2}
	\begin{algorithmic}[1]
		\State{\textbf{Input:} Measurement matrix ${\bf Y}$, dictionary ${\bf \mathring A}\in \mathbb{C}^{M \times P}$}
		\State{For given ${\bf Y}$ and ${\bf \mathring A}$, obtain $\bm{\nu}^{(0)} $ and $ {\bf X}^{(0)}$ using OMP} \label{omps}
		\State{Choose $J_{\text{max}}$ and set $j = 1$ }
		
		\While{$j \leq J_{\text{max}}$}
		\State{Set $\bm{\nu}^{(j)} = \bm{0}_{K}$ and ${\bf X}^{(j)} = \bm{0}_{L \times K}$}
		\For{$k = 1$ to $K$}
		\State{Get ${\bf Y}^{(j)}_{\text{new},k} = \sum_{k^{\prime} =1 }^{k-1} \bm{\Phi}  {\bf a}({\nu^{(j)}_{k^{\prime}} }) ({\bf x}^{(j)}_{k^{\prime}})^\mathsf{T}$}
		\State{Get ${\bf Y}^{(j)}_{\text{old},k} = \sum_{k^{\prime} = k+1}^{K} \bm{\Phi}  {\bf a}({\nu^{(j-1)}_{k^{\prime}} }) ({\bf x}^{(j-1)}_{k^{\prime}})^\mathsf{T}$   }
		
		\State{Get ${\bf Y}^{(j)}_k = {\bf Y} - {\bf Y}^{(j)}_{\text{new},k} - {\bf Y}^{(j)}_{\text{old},k}$} \label{stepYk}
		\State{Get $[{\nu}^{(j)}_k, {{\bf x}}^{(j)}_k] \leftarrow \text{GOMP}\big({\bf Y}^{(j)}_k, \nu^{(j-1)}_k, {\bf x}^{(j-1)}_k \big) $}\label{stepGOMP}
		\EndFor
		\EndWhile
		\State{\textbf{Output:} The estimated ${\bm{ \hat \nu}}$ and ${{\bf \hat X}}$}
	\end{algorithmic}
\end{algorithm}

\section{Proposed EGD for projection matrix design}\label{sec4}

From the above, it is clear that the estimation accuracy and the complexity of the proposed GOMP method (i.e., Algorithm~\ref{GOMP2}) is highly dependent on the initial on-grid estimates $\bm{\nu}^{(0)} $ and $ {\bf X}^{(0)}$. Specifically, we have observed that if the OMP method correctly approximates the true frequencies $\bm{\nu}$ to its nearest on-grid points $\bm{\nu}^{(0)}$ (i.e., $ \nu^{(0)}_k  = \underset{\mathring \nu_p \in \bm{\mathring \nu}}{\arg\min} |\nu_k -  \mathring \nu_p|, \forall k$), the refinement steps of Algorithm \ref{GOMP} always lead to a high accuracy estimation. Like any CS-based method, the performance of the OMP is dependent on the employed sensing matrix that is given by  ${\bf \Psi} = {\bf \Phi} {\bf \mathring A}(\bm{\mathring \nu}) \in \mathbb{C}^{N\times P}$, which should be designed carefully so that it satisfies a certain property, e.g., by minimizing its mutual coherence $\mu_{\max}({\bf \Psi})$ defined as 
\begin{align}\label{Mu}
\mu_{\max}({\bf \Psi}) = \max_{i \neq j} \frac{|{\boldsymbol \psi}_i^{\text{H}} {\boldsymbol \psi}_j|}{\Vert {\boldsymbol \psi}_i \Vert_2 \Vert {\boldsymbol \psi}_j \Vert_2},
\end{align}
with columns ${\boldsymbol \psi}_p \in \mathbb{C}^{N}, p \in \{1,\dots, P\}$. Clearly, a large coherence $\mu_{\max}({\bf \Psi})$ means that there exist, at least, two highly correlated columns in ${\bf \Psi}$, which may confuse any sparse coding technique like OMP. Nonetheless, if $K < \frac{1}{2}\Big(1 + \frac{1}{\mu_{\max}({\bf \Psi})} \Big)$, the OMP is guaranteed to correctly approximates $\bm{\nu}$ to the nearest on-grid points $\bm{\nu}^{(0)}$ with overwhelming probability \cite{ardah2021recovery}. 


Formally, by assuming that the dictionary ${\bf \mathring A}(\bm{\mathring \nu}) \in \mathbb{C}^{M\times P}$ is fixed, the problem of sensing matrix design reduces to finding the projection matrix ${\bf \Phi} \in \mathbb{C}^{N\times M}$ with CM entries so that the coherence $\mu_{\max}({\bf \Psi})$ is minimized, which can be expressed as 
\begin{equation}
\begin{aligned}\label{MCM1}
\underset{{\bf \Phi} }{\min}  & \quad  \mu_{\max}({\bf \Psi}) \\
\text{s.t. } & \quad \Vert { {\boldsymbol \psi}_p } \Vert_2 = 1, \forall p, \text{ and }
 |\phi_{n,m}| = 1, \forall n,m.
\end{aligned}
\end{equation}

Note that problem (\ref{MCM1}) is \kr{non} non-convex and NP-hard \cite{CSSMCM}. Therefore, we propose to solve (\ref{MCM1}) indirectly as \cite{ardah2021recovery}
\begin{equation} 
\begin{aligned}\label{MCM_F}
\underset{{\bf \Phi} }{\min} \text{ }  & \quad   \Vert  {\bf \Psi}^{\text{H}} {\bf \Psi}  - {\bf I}_{P}  \Vert^2_{\text{F}} \\
 \text{s.t. } & \quad  \Vert { {\boldsymbol \psi}_p } \Vert_2 = 1, \forall p, \text{ and }  |\phi_{n,m}| = 1, \forall n,m.
\end{aligned} 
\end{equation}
   
Similarly to \cite{GDes2}, we reformulate (\ref{MCM_F}) so that the unit-norm constraints are directly embedded into the objective function. To this end, let $\bm{\Psi} = {\bf Q} {\bf D}$, where ${\bf Q} = {\bf \Phi} {\bf \mathring A}(\bm{\mathring \nu}) = [{\bf q}_1,\dots, {\bf q}_P]\in \mathbb{C}^{N\times P}$ and ${\bf D} = \text{ diag} \Big\{\frac{1}{\Vert{\bf q}_1\Vert_2}, \dots, \frac{1}{\Vert{\bf q}_P\Vert_2} \Big\} \in \mathbb{R}^{P\times P}$. Since the columns of \kr{the} matrix $\bm{\Psi}$  have unit-norm, (\ref{MCM_F}) can be rewritten as
\begin{equation} 
\begin{aligned}\label{MCM_F2}
\underset{{\bf \Phi} }{\min} \text{ }  & \quad   \eta( {\bf \Phi}  ) \\
\text{s.t. } & \quad  {\bf \Psi}  = {\bf Q}{\bf D}  \text{ and } {\bf Q} = {\bf \Phi} {\bf \mathring A}(\bm{\mathring \nu}),\\ 
& \quad {\bf D} = \text{ diag} \Bigg\{\frac{1}{\Vert{\bf q}_1\Vert_2}, \dots, \frac{1}{\Vert{\bf q}_P\Vert_2} \Bigg\}, \\
& \quad  |\phi_{n,m}| = 1, \forall n,m,
\end{aligned} 
\end{equation}
where $\eta( {\bf \Phi}  ) = \Vert  {\bf D} {\bf Q}^{\text{H}} {\bf Q}{\bf D} - {\bf I}_{P}  \Vert^2_{\text{F}}$ and ${\bf D} = {\bf D}^{\mathsf{H}}$ due to its diagonal and real-valued structure. To obtain a solution for (\ref{MCM_F2}), we propose a CM constrained GD-based method, which updates the projection matrix ${\bf \Phi}$ iteratively as
\begin{align}\label{phit}
{\bf \Phi}^{(t)} = {\bf \Pi} \bigg( {\bf \Phi}^{(t-1)} - \zeta \cdot  \frac{\partial \eta( {\bf \Phi}  )}{\partial {\bf \Phi}} \bigg|_{{\bf \Phi} = {\bf \Phi}^{(t-1)}}  \bigg),
\end{align}
where $t$ is the iteration index, $\zeta$ is the step-size, ${\bf \Pi}(\cdot)$ is a projection function that imposes the CM constraints on the entries of the input matrix entry-wise, i.e., ${\bf \Pi}(z) = \frac{z}{|z|}$, and $\frac{\partial \eta( {\bf \Phi}  )}{\partial {\bf \Phi}}$ is the gradient of $\eta( {\bf \Phi}  )$ w.r.t ${\bf \Phi}$, which is given as \cite{GDes2}
\begin{equation}
\begin{aligned}\label{GradPhi}
\frac{\partial \eta( {\bf \Phi}  )}{\partial {\bf \Phi} }  = & \text{ } 4  {\bf Q}{\bf D}{\bf E}{\bf D} {\bf \mathring A}(\bm{\mathring \nu})^{\text{H}}   - 2  {\bf \Phi} {\bf \mathring A}(\bm{\mathring \nu}){\bf R} {\bf \mathring A}(\bm{\mathring \nu})^{\text{H}},
\end{aligned}
\end{equation}
where ${\bf E} = {\bf D} {\bf Q}^{\text{H}} {\bf Q}{\bf D} - {\bf I}_{P}$, ${\bf R} = \text{diag}\{ [{\bf C}]_{[1,1]}, \dots, [{\bf C}]_{[P,P]} \}$, and ${\bf C} = 2{\bf E} {\bf D} {\bf Q}^{\text{H}} {\bf Q} {\bf D}^{3} $. Note that, the update step in (\ref{phit}) is a direct extension of the proposed unconstrained GD method in \cite{GDes2} to accounts for the CM constraints. Our results show that both the unconstrained method of \cite{GDes2} and its direct extension in (\ref{phit}) achieve a mutual coherence that is far from the known theoretical Welch lower-bound $\beta = \sqrt{\frac{P-N}{N(P-1)}}$ \cite{CS}. To enhance their performance, we first note that ${\bf E}$ in (\ref{GradPhi}) represents the error matrix of the objective function $\eta( {\bf \Phi}  )$. Here, we propose to apply a shrinking operator on ${\bf E}$  element-wise to get $\underline{{\bf E}}$, where the $(k,j)$th entry of $\underline{{\bf E}}$, i.e., $\underline{e}_{k,j}$ is obtained as \cite{ardah2021recovery}		
\begin{align}\label{ShrinkE}
\underline{e}_{k,j}  = \begin{cases}
0, & \big|	{e}_{k,j} \big| < \alpha\cdot {\beta}  \\
\frac{{e}_{k,j}}{|{e}_{k,j}|} \cdot \Big( \big|	{e}_{k,j} \big| -  \alpha\cdot {\beta} \Big), & \text{otherwise},
\end{cases}
\end{align}
where ${e}_{k,j}$ is the $(k,j)$th entry of ${{\bf E}}$, $\alpha\geq 1$ is a relaxation parameter, and $\beta$ is the Welch lower-bound defined above. In \cite{ardah2021recovery}, we have shown that the choice of $\alpha$ highly impacts the mutual coherence $\mu_{\max}(\bm{\Psi})$ of the obtained sensing matrix, where a very-large or a very-small value leads to a high $\mu_{\max}(\bm{\Psi})$. Therefore, $\alpha$ should be chosen carefully depending on the given dictionary matrix.    

After a closer look at (\ref{ShrinkE}), we can observe that for a very tight threshold $\bar{\beta} = \alpha\cdot {\beta}$, the resulting error matrix $\underline{{\bf E}}$ becomes a sparse matrix, where some of its entries that are smaller than $\bar\beta$ will be set to zero. The direct implication of such \kr{a} shrinking operator is that the new projection matrix ${\bf \Phi}^{(t)}$ will be updated so that it mainly \kr{reduces} the entries that are larger than $\bar\beta$. In summary, the proposed GD method for sensing/projection matrix design is given by Algorithm \ref{EnhancedGrad}. 

\begin{algorithm}[t]
	\caption{Proposed GD method for sensing matrix design.}
	\label{EnhancedGrad}
	 
		\begin{algorithmic}[1]
			\State{\textbf{Inputs}: Initial ${\boldsymbol \Phi}^{(0)} \in \mathbb{C}^{N\times M}$ and dictionary ${\bf \mathring A}(\bm{\mathring \nu}) $. }
		   \State{Select $T_{\max}$ and step-size $ \zeta$. Set $t = 1$}
			\While{$t \leq T_{\max}$}
			\State{For given ${\bf \Phi}^{(t-1)}$, compute  ${\bf E}^{(t-1)}$}
			
			\State{Apply the shrinking (\ref{ShrinkE}) on ${\bf E}^{(t-1)} $ to get $\underline{{\bf E}}^{(t-1)}$}

            \State{For given $\underline{{\bf E}}^{(t-1)}$ and ${\bf \Phi}^{(t-1)}$, compute ${\bf \Phi}^{(t)}$ using (\ref{phit}) }
			
			
			\EndWhile
			\State{\textbf{Output:} Sensing matrix ${\bf \Psi}^{\star} = {\bf \Phi}^{\star} {\bf \mathring A}(\bm{\mathring \nu})$}
		\end{algorithmic}
	
\end{algorithm}

\section{Numerical Results}\label{Sec53}

\textbf{Example 1: Evaluation of Algorithm \ref{EnhancedGrad}.}
In this example, we set $N = 16$, $M = 64$, and design the dictionary matrix as ${\bf \mathring A}(\bm{\mathring \nu}) = [{\bf a}(\mathring \nu_1), \dots, {\bf a}(\mathring\nu_P)] \in \mathbb{C}^{M\times P}$, where $\mathring \nu_p = \frac{2\pi (p-1)}{P}, p\in \{1,\dots,P\}$, so that ${\bf \mathring A}(\bm{\mathring \nu}) {\bf \mathring A}(\bm{\mathring \nu})^{\mathsf{H}} = P{\bf I}_{M} $. For comparison, we include the results when the projection matrix ${\boldsymbol \Phi}$ is designed from a DFT matrix\footnote{Let ${\bf W}_{64}$ denotes the $64\times 64$ DFT matrix. Then, the projection matrix is given as ${\boldsymbol \Phi} = [{\bf W}_{64}]_{[{\bf n}, :]} \in \mathbb{C}^{16\times 64}$, where ${\bf n} = [1,5,9,13,17,\dots,64]^{\mathsf{T}}$.}, randomly\footnote{The $(i,j)$th entry of ${\boldsymbol \Phi}$ is given as $[{\boldsymbol \Phi}]_{[i,j]} = e^{j\theta_{i,j}}, \theta_{i,j} \in [0,2\pi]$.}, using the GD-based method of \cite{ardah2021recovery}, \kr{or} using the extended GD-based method of \cite{GDes2} to accounts for the CM constraints\footnote{Note that the main difference between our proposed GD-based method in Algorithm \ref{EnhancedGrad} and the GD-based method in \cite{ardah2021recovery} is that the former imposes the unit-norm constraints directly into the objective function, while the latter imposes the unit-norm constraints after updating the projection matrix on every iteration. On the other hand, the main difference between Algorithm \ref{EnhancedGrad} and the extended GD-based method of \cite{GDes2} is that the former uses the error matrix $\underline{{\bf E}}$ obtained using (\ref{ShrinkE}) when updating ${\boldsymbol \Phi}$ on every iteration, while the latter uses the error matrix ${\bf E}$ when updating ${\boldsymbol \Phi}$ on every iteration.}. For the GD-based methods, we assume that the initial projection matrix, i.e., ${\boldsymbol \Phi}^{(0)}$ is obtained using the closed-form approach discussed in \cite{ardah2021recovery}, since we have observed that it always leads to a better result than \kr{a} random initialization approach. Moreover, the relaxation parameter $\alpha$ is chosen as the one leading to the lowest coherence among several tested candidates.      

\begin{figure}[t]
	\centering
	\includegraphics[width=0.7\linewidth]{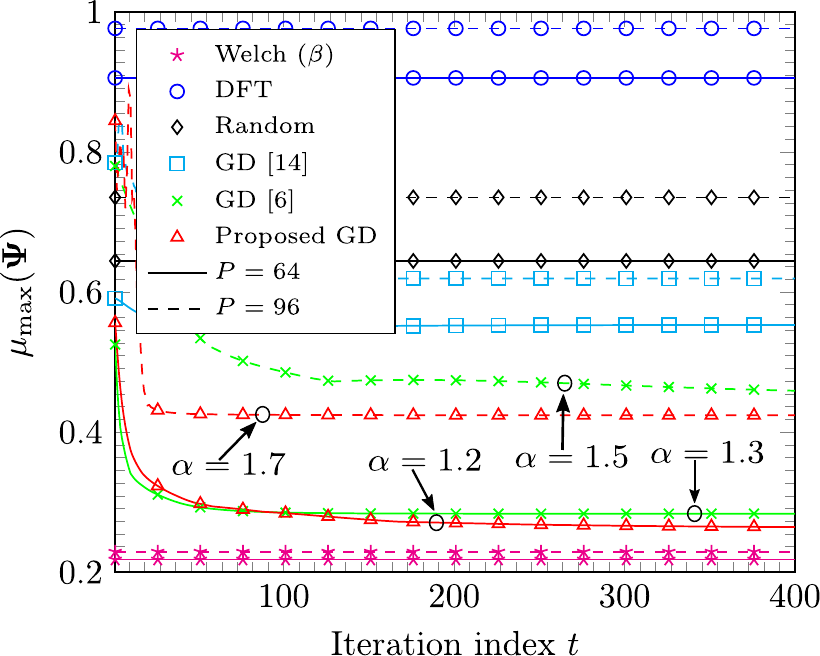}
	\vspace{-5pt}
	\caption{Mutual coherence $\mu_{\text{max}}(\bm{\Psi})$ versus iteration index $t$.}
	\label{fig1}
	\vspace{-5pt}
\end{figure}

\begin{figure}[t]
	\centering
	\includegraphics[width=1\linewidth]{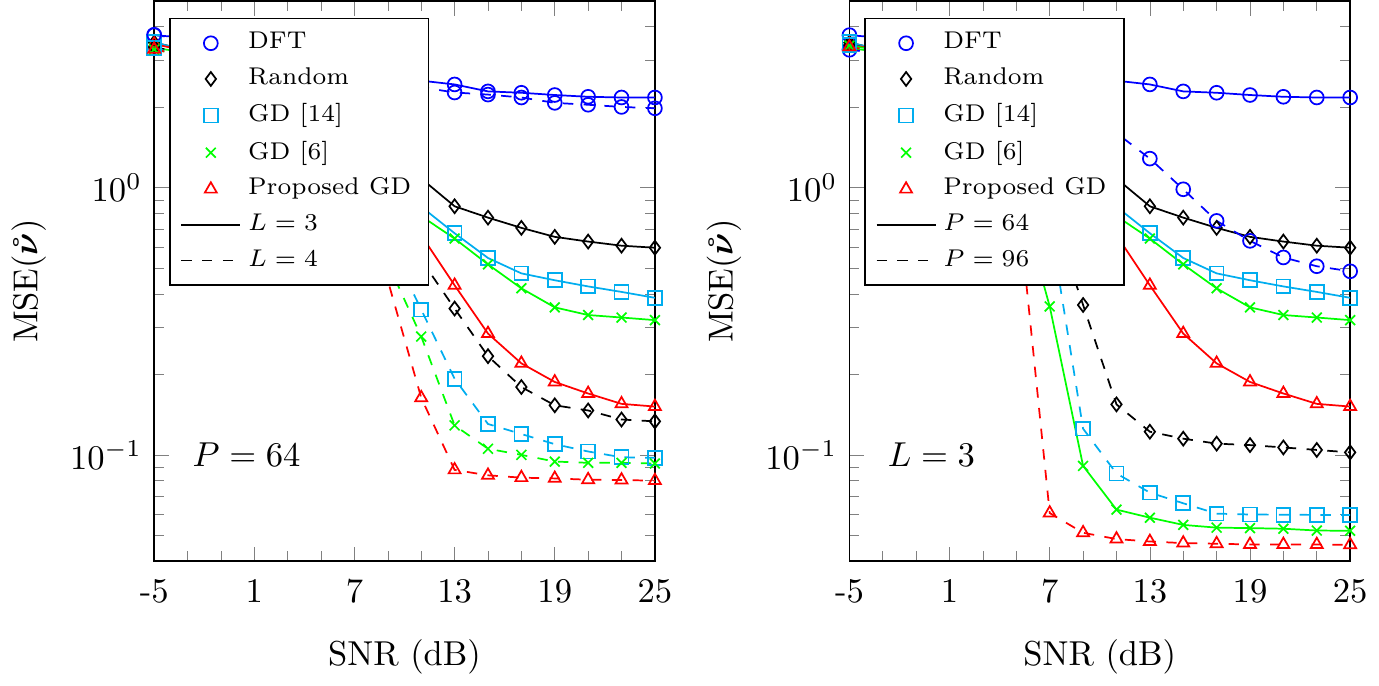}
	\vspace{-15pt}
	\caption{MSE versus SNR [$K = 5$, $N = 16$, $M = 64$].}
	\label{fig2}
	\vspace{-15pt}
\end{figure}

Fig. \ref{fig1} shows the mutual coherence $\mu_{\text{max}}(\bm{\Psi})$ versus the iteration index $t$ results for different values of $P$. It can be seen that the proposed GD-based method has the best performance in all of the considered scenarios. This shows the effectiveness of the proposed shrinking operator (\ref{ShrinkE}) on the projection matrix design\footnote{It is worth mentioning that the performance gain of the proposed GD-based method over the other benchmark methods holds true even in the unconstrained projection matrix design, i.e., when we ignore the CM constraints. The simulation results are not included here due to space limitations.}, when compared to the GD-based method of \cite{GDes2} and the effectiveness of embedding the unit-norm constraints into the objective function, when compared to the GD-based method of \cite{ardah2021recovery}. To investigate the impact of the projection matrix deign on the DoA estimation accuracy, Fig.~\ref{fig2} shows the mean-squared errors (MSE) versus the signal-to-noise-ratio (SNR) results, which are defined as $\text{MSE}(\bm{\mathring \nu}) = \mathbb{E}\{ \sum_k | \nu_k -  \mathring \nu_k |^2 \}$  and $\text{SNR} = \mathbb{E} \{ { \Vert {\bf Y} - {\bf N} \Vert_{\text{F}}^2 }\} / \mathbb{E} \{ {\Vert {\bf N}  \Vert^2_{\text{F}}} \}$ (in dB). Here, we assume that the OMP method \cite{OMPComp} uses the dictionary matrix ${\bf \mathring A} \in \mathbb{C}^{M\times P}$ to approximate $\bm{\nu} = [\nu_1,\dots,\nu_K]$ with $\nu_k \in [0,2\pi], \forall k$, to the best on-grid points $\bm{ \mathring \nu} = [\mathring \nu_1,\dots, \mathring\nu_K]$. The figure clearly shows that the accuracy of the OMP can be significantly improved by designing a projection matrix with a low mutual coherence $\mu_{\max}\{{\bf \Psi}\}$, especially with smaller number of time samples, $L$, and dictionary grid points, $P$. 

\textbf{Example 2: Evaluation of Algorithm \ref{GOMP2}.} Here, we show simulation results comparing our proposed GOMP method with two high-resolution DoA estimation methods: CS-based Newtonized OMP (NOMP) \cite{NOMP} and DFT beamspace ESPRIT (BS-ESPRIT) \cite{StESBRIT}. To comply with the BS-ESPRIT requirements as discussed in \cite[Lemma 1]{StESBRIT}, we assume that $\nu_k \in [0,  \nu_{\max} ], \forall k$, where ${ \nu_{\max}} = \frac{2\pi (N-1)}{M}$. For CS-based OMP, NOMP, and GOMP methods, we assume that the proposed GD-based method is used to design the projection matrix ${\bf \Phi}$ for a given dictionary matrix that is formed as ${\bf \mathring A}(\bm{\mathring \nu}) = [{\bf a}(\mathring \nu_1), \dots, {\bf a}(\mathring\nu_P)] \in \mathbb{C}^{M\times P}$ with $\mathring \nu_p = \frac{ \nu_{\max} (p-1)}{P}, p\in \{1,\dots,P\}$. Note that in this case, ${\bf \mathring A}(\bm{\mathring \nu}) {\bf \mathring A}(\bm{\mathring \nu})^{\mathsf{H}} \neq P{\bf I}_{M} $. Finally, for NOMP and GOMP, we set the maximum number of iterations as $I_{\text{max}} = 10$ and $ J_{\text{max}} = 5$.

\begin{figure}[t]
	\centering
	\includegraphics[width=1\linewidth]{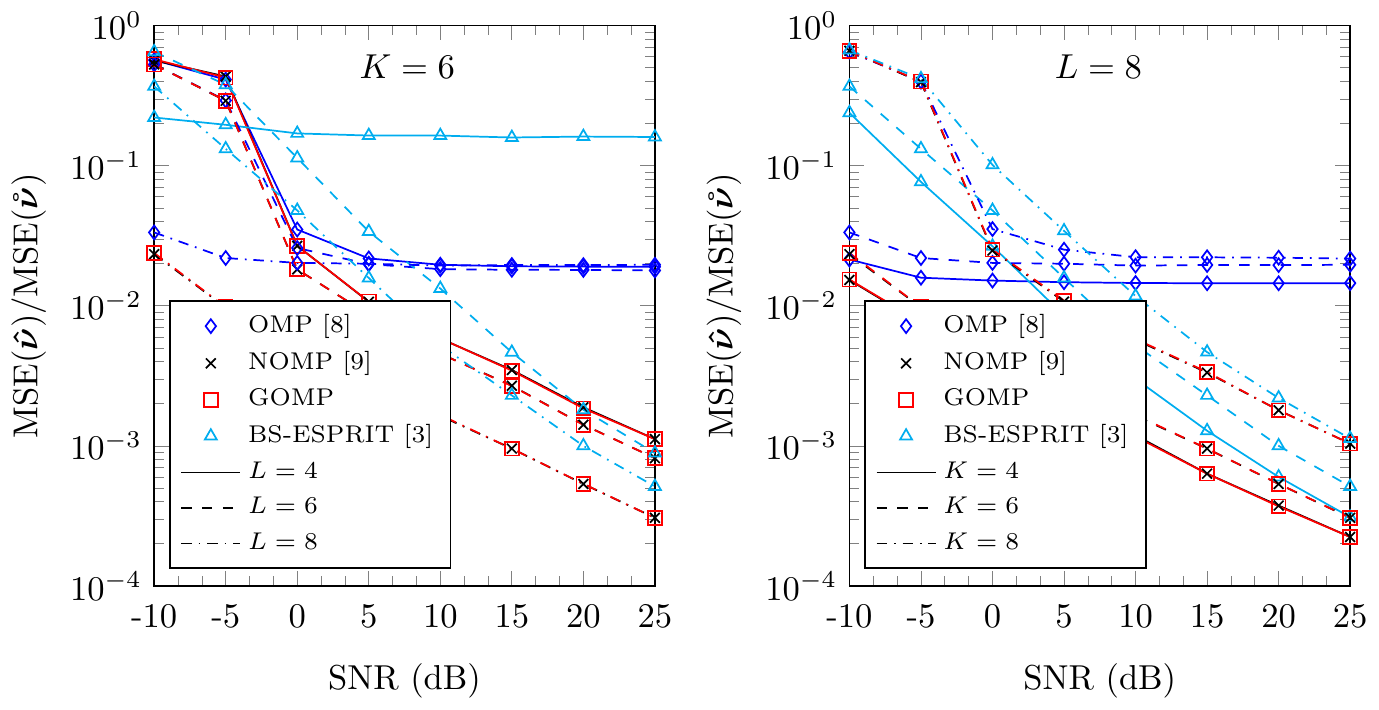}
	\vspace{-15pt}
	\caption{MSE versus SNR assuming $N = 16$, $M = 64$, $P = 64$.}
	\label{fig3}
	\vspace{-15pt}
\end{figure}

Fig. \ref{fig3} shows MSE($\bm{\hat \nu}$) and $\text{MSE}(\bm{ \mathring \nu})$ versus SNR results for various system settings, where $\bm{\hat \nu}$ is the high-resolution estimated parameters vector and MSE($\bm{\hat \nu}$) is defined similarly to $\text{MSE}(\bm{\mathring \nu})$. From the figure, we can see that the CS-based methods have better DoA estimation accuracy than the subspace-based BS-ESPRIT method, especially  in the low SNR regime. Note that, BS-ESPRIT requires a sufficient number of time samples, i.e.,  $L\geq K$ to have an accurate DoA estimation. On the other hand, this condition is not required by the CS-based methods, although in the expense of higher computational complexity. Moreover, Fig. \ref{fig3} shows that GOMP, in average, has a similar estimation accuracy compared to NOMP. Nonetheless, in our simulations, we noted that in some realizations, GOMP has higher accuracy than NOMP and vice-versa on the other realizations.

\vspace{-5pt}

\section{Conclusions}

\vspace{-5pt}

In this paper, we have proposed the GOMP method for high-resolution DoA estimation and a new GD-based method for projection/sensing matrix design while taking into account the constant modulus constraints imposed by the projection matrix hardware components. Simulation results are provided showing the effectiveness of the proposed methods.

\bibliographystyle{IEEEtran}
\bibliography{refs}

\begin{thebibliography}{10}
\providecommand{\url}[1]{#1}
\csname url@samestyle\endcsname
\providecommand{\newblock}{\relax}
\providecommand{\bibinfo}[2]{#2}
\providecommand{\BIBentrySTDinterwordspacing}{\spaceskip=0pt\relax}
\providecommand{\BIBentryALTinterwordstretchfactor}{4}
\providecommand{\BIBentryALTinterwordspacing}{\spaceskip=\fontdimen2\font plus
\BIBentryALTinterwordstretchfactor\fontdimen3\font minus
  \fontdimen4\font\relax}
\providecommand{\BIBforeignlanguage}[2]{{%
\expandafter\ifx\csname l@#1\endcsname\relax
\typeout{** WARNING: IEEEtran.bst: No hyphenation pattern has been}%
\typeout{** loaded for the language `#1'. Using the pattern for}%
\typeout{** the default language instead.}%
\else
\language=\csname l@#1\endcsname
\fi
#2}}
\providecommand{\BIBdecl}{\relax}
\BIBdecl

\bibitem{MLDoA}
P.~Stoica and K.~Sharman, ``Maximum likelihood methods for direction-of-arrival
  estimation,'' \emph{IEEE Transactions on Acoustics, Speech, and Signal
  Processing}, vol.~38, no.~7, pp. 1132--1143, 1990.

\bibitem{MUSIC}
P.~Gupta and S.~Kar, ``{MUSIC} and improved {MUSIC} algorithm to estimate
  direction of arrival,'' in \emph{Proc. International Conference on
  Communications and Signal Processing (ICCSP)}, 2015, pp. 0757--0761.

\bibitem{StESBRIT}
J.~{Zhang} and M.~{Haardt}, ``Channel estimation and training design for hybrid
  multi-carrier mmwave massive {MIMO} systems: The beamspace {ESPRIT}
  approach,'' in \emph{Proc. 25th European Signal Processing Conference
  (EUSIPCO)}, Sep. 2017, pp. 385--389.

\bibitem{CS}
D.~L. {Donoho}, ``Compressed sensing,'' \emph{IEEE Transactions on Information
  Theory}, vol.~52, no.~4, pp. 1289--1306, Apr. 2006.

\bibitem{CSBounds}
H.~{Reboredo}, F.~{Renna}, R.~{Calderbank}, and M.~R.~D. {Rodrigues}, ``Bounds
  on the number of measurements for reliable compressive classification,''
  \emph{IEEE Trans. Signal Process.}, vol.~64, no.~22, pp. 5778--5793, Aug.
  2016.

\bibitem{ardah2021recovery}
K.~Ardah, M.~Haardt, T.~Liu, F.~Matter, M.~Pesavento, and M.~E. Pfetsch,
  ``Recovery under side constraints,'' in \emph{Compressed Sensing in
  Information Processing}, G.~Kutyniok, H.~Rauhut, and R.~J. Kunsch, Eds.\hskip
  1em plus 0.5em minus 0.4em\relax Switzerland: Springer Nature, 2021,
  published in the book series {Applied and Numerical Harmonic Analysis}.

\bibitem{Boche2015}
H.~Boche, R.~Calderbank, G.~Kutyniok, and J.~Vyb{\'i}ral, \emph{A Survey of
  Compressed Sensing}.\hskip 1em plus 0.5em minus 0.4em\relax Cham: Springer
  International Publishing, 2015, pp. 1--39.

\bibitem{OMPComp}
B.~L. {Sturm} and M.~G. {Christensen}, ``Comparison of orthogonal matching
  pursuit implementations,'' in \emph{Proc. of the 20th European Signal
  Processing Conference (EUSIPCO)}, Aug. 2012, pp. 220--224.

\bibitem{NOMP}
B.~{Mamandipoor}, D.~{Ramasamy}, and U.~{Madhow}, ``Newtonized orthogonal
  matching pursuit: Frequency estimation over the continuum,'' \emph{IEEE
  Transactions on Signal Processing}, vol.~64, no.~19, pp. 5066--5081, Jun.
  2016.

\bibitem{ardah_icassp19}
K.~{Ardah}, A.~L.~F. {de Almeida}, and M.~{Haardt}, ``A gridless {CS} approach
  for channel estimation in hybrid massive {MIMO} systems,'' in \emph{Proc.
  IEEE International Conference on Acoustics, Speech and Signal Processing
  (ICASSP)}, May 2019, pp. 4160--4164.

\bibitem{ardah_icassp2020}
K.~{Ardah}, B.~{Sokal}, A.~L.~F. {de Almeida}, and M.~{Haardt}, ``Compressed
  sensing based channel estimation and open-loop training design for hybrid
  analog-digital massive {MIMO} systems,'' in \emph{Proc. IEEE International
  Conference on Acoustics, Speech and Signal Processing (ICASSP)}, May 2020,
  pp. 4597--4601.

\bibitem{CSSMCM}
K.~{Ardah}, M.~{Pesavento}, and M.~{Haardt}, ``A novel sensing matrix design
  for compressed sensing via mutual coherence minimization,'' in \emph{Proc.
  IEEE 8th International Workshop on Computational Advances in Multi-Sensor
  Adaptive Processing (CAMSAP)}, Dec. 2019, pp. 66--70.

\bibitem{GDes}
V.~{Abolghasemi}, S.~{Ferdowsi}, B.~{Makkiabadi}, and S.~{Sanei}, ``On
  optimization of the measurement matrix for compressive sensing,'' in
  \emph{Proc. 18th European Signal Processing Conference}, Aug. 2010, pp.
  427--431.

\bibitem{GDes2}
H.~Bai, C.~Hong, and X.~Li, ``Construction of unit-norm tight frame based
  preconditioner for sparse coding,'' in \emph{Proc. IEEE International
  Conference on Acoustics, Speech and Signal Processing (ICASSP)}, 2021, pp.
  5420--5424.

\bibitem{Direct}
C.~Lu, H.~Li, and Z.~Lin, ``Optimized projections for compressed sensing via
  direct mutual coherence minimization,'' \emph{Signal Processing}, vol. 151,
  pp. 45 -- 55, 2018.

\bibitem{SVDShrinkage}
L.~{Yu}, G.~{Li}, and L.~{Chang}, ``Optimizing projection matrix for compressed
  sensing systems,'' in \emph{Proc. 8th International Conference on
  Information, Communications Signal Processing (ICICS)}, 2011, pp. 1--5.

\bibitem{ardah_unify}
K.~{Ardah}, G.~{Fodor}, Y.~C.~B. {Silva}, W.~C. {Freitas}, and F.~R.~P.
  {Cavalcanti}, ``A unifying design of hybrid beamforming architectures
  employing phase shifters or switches,'' \emph{IEEE Trans. Veh. Technol.},
  vol.~67, no.~11, pp. 11\,243--11\,247, Nov. 2018.

\end{thebibliography}
		
\end{document}